\documentclass[10pt,prb,aps,twocolumn,showpacs,amsmath,amssymb]{revtex4-1}
\usepackage{times}
\usepackage{textcomp}
\usepackage{graphicx}
\usepackage{bm}
\allowdisplaybreaks 
\usepackage{braket}
\usepackage[breaklinks]{hyperref}
\hypersetup{colorlinks=true, linkcolor=blue, citecolor=blue, filecolor=blue, urlcolor=blue}

\begin{document}

\title{Symmetry breaking and fermionic fractional Chern insulator in topologically trivial bands}

\author{Stefanos Kourtis}

\affiliation{Department of Physics, Boston University, Boston, MA, 02215, USA}

\date{\rm\today}

\begin{abstract}
We describe a mechanism by which fermions in topologically trivial bands can form correlated states exhibiting a fractional quantum Hall (FQH) effect upon introduction of strong repulsive interactions. These states are solid-liquid composites, in which a FQH liquid is induced by the formation of charge order (CO), following a recently proposed paradigm of symmetry-breaking topological (SBT) order [Phys.~Rev.~Lett.~{\bf 113}, 216404 (2014)]. We devise a spinless fermion model on a triangular lattice, featuring a topologically trivial phase when interactions are omitted. Adding strong short-range repulsion, we first establish a repulsion-driven CO phase at density $\rho_{\mathrm{CO}}=2/3$ particles per site, then dope the model to higher densities $\rho = \rho_{\mathrm{CO}} + \nu/6$. At $\nu=1/3,2/5$ ($\rho=13/18,11/15$), we observe definitive signatures of both CO and the FQH effect --- sharply peaked static structure factor, gapped and degenerate energy spectrum and fractionally quantized Hall conductivity $\sigma_{\mathrm{H}}=1/3,2/5$ in units of $e^2/h$ --- over a range of all model parameters. We thus obtain direct evidence for fermionic SBT order of FQH type in topologically trivial bands.
\end{abstract}

\maketitle

\section{Introduction}

The discovery of the fractional quantum Hall (FQH) effect~\cite{Tsui1982} and its interpretation in terms of topological order and fractionally charged quasiparticles with anyonic statistics~\cite{Laughlin1983a,Halperin1984} have led to a large body of research on the theoretical underpinnings of FQH states, as well as their potential to power functionality that goes far beyond conventional electronics~\cite{Nayak2008}. The identification of the quantized Hall conductivity as a topological invariant, known as the Chern number, in the presence of an underlying periodic lattice, disorder, and interactions~\cite{Thouless1982,Kohmoto1985,Niu1985a} and in the absence of a net magnetic field~\cite{Haldane1988}, ushered in the era of topological materials and multiplied the number of candidate hosts for FQH phases.

FQH-type topological order is a result of correlations, and as such it is anticipated to be intricately related to interaction-driven symmetry breaking. This is particularly pertinent in lattice systems, where geometric effects of interactions are pronounced. In this context, symmetry breaking can be influential on many levels. First, lattice models that harbor FQH states~\cite{Kliros1991}, called fractional Chern insulators (FCI)~\cite{Neupert2011,Sheng2011,Regnault2011}, require that time reversal (TR) symmetry is broken, as this is a prerequisite for bands with nonzero Chern number, or simply Chern bands. These arise in a variety of physical settings, such as optical lattices with artificial gauge fields~\cite{Jotzu2014,Aidelsburger2014,Nascimbene2015a}, layered materials and heterostructures~\cite{Xiao2011,Grushin2013,Chang2013,Trier2016}, and itinerant magnets in frustrated lattices~\cite{Ohgushi2000,Taguchi2001,Martin2008}. Within Chern bands, correlated states that either compete or coexist with topological order can emerge with further breaking of symmetries~\cite{Neupert2011a,Kourtis2013,Kumar2014a,Chen2015,Doretto2015}. On the other hand, TR-symmetric versions of FCI states may arise in topologically nontrivial bands characterized by a nonzero $\mathbb{Z}_2$ invariant instead of a Chern number~\cite{Levin2009,Neupert2011a,Repellin2014}.

A tantalizing prospect for FCI topological order is to forgo topologically nontrivial bands altogether: topological order could emerge spontaneously, driven solely by the interactions between itinerant particles. This idea inspired the search for a topological Mott insulator~\cite{Raghu2008}. Unfortunately, thorough investigations beyond the mean-field level showed that, in the originally proposed context, the topological Mott insulating phase is unstable against charge ordering~\cite{Capponi2017}. On the other hand, compelling evidence for interaction-driven spontaneous TR symmetry breaking~\cite{Tieleman2012}, fermionic \textit{integer}~\cite{Tieleman2012,Zhu2016a}, and bosonic \textit{fractional} quantum Hall effect~\cite{Gong2015,Zhu2016b} on frustrated lattices have rekindled interest in this prospect.

On the other hand, \textit{fermionic} FCI states that arise spontaneously in topologically trivial bands have been elusive. Off-diagonal interactions~\cite{Bukov2015,Kourtis2015,Simon2015a}, which appear in periodically driven cold atoms as a secondary effect, can favor fermionic FCIs without a nonzero single-particle Berry curvature~\cite{Kourtis2015}, but a physically relevant setting where such interactions are dominant is lacking. A more natural platform is that of multi-orbital models of strongly correlated electrons on frustrated lattices~\cite{Venderbos2011,Venderbos2011a,Kourtis2012a}. Here interactions induce noncoplanar spin ordering that breaks translation and TR symmetries and generates a Chern band for \textit{part} of the electron density, much like in a frustrated Kondo lattice model~\cite{Ohgushi2000,Taguchi2001,Martin2008}, even though without interactions the system is topologically trivial. When doped to a FCI filling fraction, such an emergent Chern band can host FCI states. Although physically viable, this scenario is challenging to address entirely on the same footing, due to the multitude of interacting degrees of freedom. Due to this, the analysis is typically split into (a) showing that a Chern band arises spontaneously, and (b) writing an effective model for this Chern band and showing that it yields FCI states upon introduction of repulsion when fractionally filled.

This work presents a mechanism by which a strongly correlated FCI state can be induced by adding strong short-range repulsion in topologically trivial bands of spinless fermions. We draw inspiration from supersolidity in the triangular lattice~\cite{Wessel2005,Boninsegni2005,Heidarian2005,Melko2005,Burkov2005}, as well as its fermionic counterpart~\cite{Hotta2006,Nishimoto2008a,Morohoshi2008,Nishimoto2009,Cano-Cortes2011a}, where the particle density spontaneously splits to solid and liquid components due to strong repulsive interactions. This mechanism was previously shown to give rise to exotic topological order in a model of nontrivial Chern bands~\cite{Kourtis2013}. Here we use it as a tool to inform the design of a lattice model, whose noninteracting part breaks time reversal symmetry explicitly and exhibits both topologically nontrivial phases with Chern numbers $C_\pm \not= 0$ and a trivial phase with $C_\pm=0$. By introducing strong short-range repulsion deep inside the trivial phase, we show that the system forms a solid-FCI composite, thus realizing symmetry-breaking topological (SBT) order. This is the first proof-of-principle example of FQH-like topological order arising upon inclusion of short-range repulsion in a minimal model of two topologically trivial bands.

We detail the minimal triangular-lattice spinless-fermion model and its phase diagram, which contains a topologically trivial $C_\pm=0$ phase, in Sec.~\ref{sec:model}. We then introduce strong short-range repulsion in the trivial phase and study the fate of the many-body system with exact diagonalization. In Sec.~\ref{sec:co} we present the signatures of robust charge order at density $\rho=2/3$ in the strong nearest-neighbor repulsion limit, on which we focus. In Sec.~\ref{sec:sbt} we present compelling evidence for FQH-type SBT order at densities $\rho=13/18,11/15$ upon inclusion of second- and third-neighbor repulsion. In Sec.~\ref{sec:ps} we investigate in detail the effects of interactions and observe the stability of an extended SBT phase against phase separation. We conclude in Sec.~\ref{sec:outlook} with a short outlook.

\section{Model}\label{sec:model}

\begin{figure}[t]
\includegraphics[width=0.99\columnwidth]{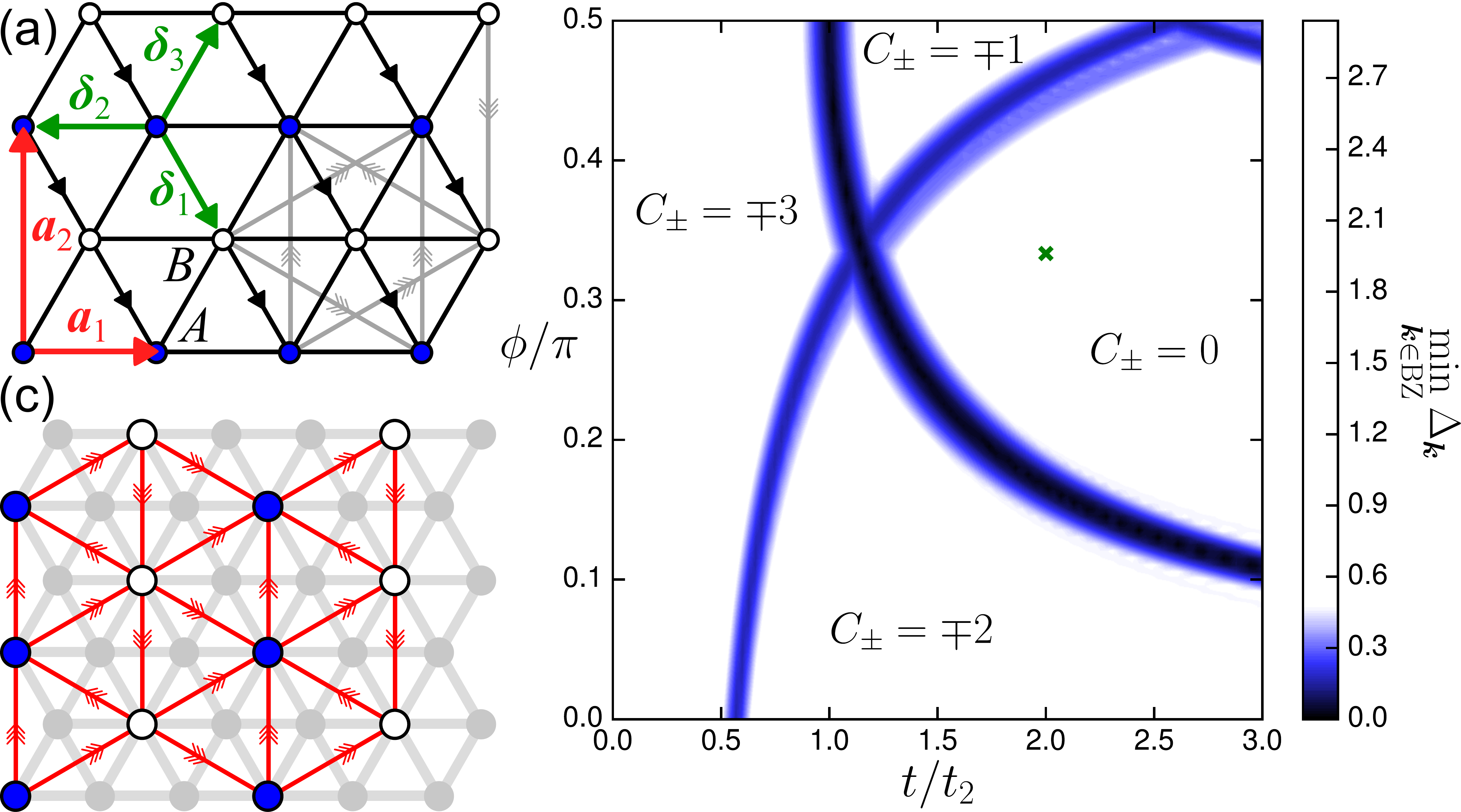}
\caption{(a) Schematic of the lattice and hopping terms of Eq.~\eqref{eq:model}. Black (gray) lines denote (next-)NN hoppings. Hopping in the direction of an arrow adds $\phi$ ($\pi/2$) to the electron wave function. Not all second NN hoppings are shown. (b) Phase diagram of the noninteracting part of model Eq.~\eqref{eq:model}. The color scale represents the minimum of the energy gap $\Delta_{\bm{k}} = \varepsilon_{\bm{k}+} - \varepsilon_{\bm{k}-}\,$. (c) Illustration of $\rho=2/3$ CO and residual triangular lattice. Occupied sites and blocked NN bonds are drawn as gray full circles and thick gray lines, respectively, and the dotted line denotes the unit cell of the effective model of dopants discussed in Sec.~\ref{sec:sbt}. The results presented below are for the parameter choice marked by ``$\times$''.}
\label{fig:model}
\end{figure}

Consider a two-dimensional system of $N$ spinless fermions on a lattice of $L=L^{\,}_{1} \times L^{\,}_{2}$ unit cells, with primitive translation vectors $\bm{a}^{\,}_{1}=(1,0)^{\mathsf{T}}$ and $\bm{a}^{\,}_{2}=(0,\sqrt{3})^{\mathsf{T}}$ and sublattices $A$ and $B$. The Hamiltonian is
\begin{subequations}
\begin{equation}
\widehat{H}= \widehat{H}^{\,}_{\textrm{kin}} + \widehat{H}^{\,}_{\textrm{int}}.
\end{equation}
The kinetic term $\widehat{H}^{\,}_{\textrm{kin}}$ is written in reciprocal space as
\begin{equation}
\widehat{H}^{\,}_{\textrm{kin}} = \sum_{{\bm{k}}\in\textrm{BZ}} \widehat{\Psi}^{\dag}_{{\bm{k}}}\,\mathcal{H}^{\,}_{\bm{k}}\,\widehat{\Psi}^{\,}_{\bm{k}},
\end{equation} 
where $\widehat{\Psi}^{\dag}_{{\bm{k}}}\equiv (\widehat{c}^{\dag}_{{\bm{k}},A}\,,\widehat{c}^{\dag}_{{\bm{k}},B})$ is the spinor of creation operators for a fermion with wavevector $\bm{k}$ in the first Brillouin zone (BZ) and sublattice index $A$ and $B$. The $2\times 2$ matrix $\mathcal{H}^{\,}_{{\bm{k}}}$ is
\begin{equation}
\mathcal{H}^{\,}_{{\bm{k}}} = (d^{\,}_{0,\bm{k}} + \mu)\,\tau^{\,}_{0} + \bm{d}^{\,}_{\bm{k}}\cdot\bm{\tau} \,,
\end{equation}
where $\tau^{\,}_{0}$ and $\bm{\tau}=(\tau^{\,}_{1},\tau^{\,}_{2},\tau^{\,}_{3})$ are the $2\times2$ unit and Pauli matrices in sublattice space, respectively, and
\begin{align}
 d^{\,}_{0,\bm{k}} =&{\ } 2t \cos \bm{k}\cdot\boldsymbol\delta_2 \,,\\
 d^{\,}_{1,\bm{k}} =&{\ } 2t [ \cos(\bm{k}\cdot\boldsymbol\delta_1 + \phi) + \cos \bm{k}\cdot\boldsymbol\delta_3 ] + 2 t_2 \sin \bm{k}\cdot\boldsymbol\zeta_2 \,,\\
 d^{\,}_{2,\bm{k}} =&{\ } - 2 t_2 \cos \bm{k}\cdot\boldsymbol\zeta_3 \,,\\
 d^{\,}_{3,\bm{k}} =&{\ } - 2 t_2 \sin \bm{k}\cdot\boldsymbol\zeta_1 \,,
\end{align}
where $t$ and $t_2$ are the amplitudes for hopping between nearest-neighbor (NN) and next NN pairs, $\mu$ is a chemical potential, $\bm\delta^{\,}_{1}=(1/2,-\sqrt{3}/2)^{\mathsf{T}}$, $\bm\delta^{\,}_{2}=(-1,0)^{\mathsf{T}}$, $\bm\delta^{\,}_{3}= -(\bm\delta^{\,}_{1}+\bm\delta^{\,}_{2})$, $\boldsymbol\zeta_1 = \boldsymbol\delta_3 - \boldsymbol\delta_1$, $\boldsymbol\zeta_2 = \boldsymbol\delta_3 - \boldsymbol\delta_2$, and $\boldsymbol\zeta_3 = \boldsymbol\delta_1 - \boldsymbol\delta_2$. The energy bands are $\varepsilon_{\bm{k}\pm}=d^{\,}_{0,\bm{k}} \pm d_{\bm{k}}$, where $d_{\bm{k}}=|\bm{d}_{\bm{k}}|$. The kinetic terms of the model are sketched in Fig.~\ref{fig:model}(a). The interactions are defined as
\begin{align}
\widehat{H}^{\,}_{\textrm{int}} =&{\ } V_1 \sum_{\langle \bm{i},\bm{j} \rangle} \widehat{n}_{\bm{i}} \widehat{n}_{\bm{j}} + V_2 \sum_{\langle\langle \bm{i},\bm{j} \rangle\rangle} \widehat{n}_{\bm{i}} \widehat{n}_{\bm{j}} + V_3 \sum_{\langle\langle\langle \bm{i},\bm{j} \rangle\rangle\rangle} \widehat{n}_{\bm{i}} \widehat{n}_{\bm{j}}  \,,
\end{align}\label{eq:model}%
\end{subequations}
where $\widehat{n}_{\bm{i}} = \widehat{c}_{\bm{i}}^\dagger \widehat{c}_{\bm{i}}$ is the fermion counting operator at lattice position $\bm{i}$, and repulsion between first-, second-, and third-nearest neighboring site pairs --- denoted as $\langle \bm{i},\bm{j} \rangle$, $\langle\langle \bm{i},\bm{j} \rangle\rangle$, and $\langle\langle\langle \bm{i},\bm{j} \rangle\rangle\rangle$, respectively --- has strength $V_1$, $V_2$, and $V_3$.

Without interactions ($V_1=V_2=V_3=0$), the phase diagram as a function of the remaining two free parameters $t_2$ and $\phi$ is shown in Fig.~\ref{fig:model}(b). In the range $t/t_2 \in [0,3]$ and $\phi/\pi \in [0,\pi/2]$ there are four phases, labeled by the Chern number that characterizes the wavefunction associated with each of the two bands,
\begin{equation}
 C_\pm = \pm \frac{1}{2\pi} \oint_{\mathrm{BZ}} \mathrm{d}\bm{k} \, \frac{\bm{d}_{\bm{k}}}{2d_{\bm{k}}^3} \cdot ( \partial_x \bm{d}_{\bm{k}} \times \partial_y \bm{d}_{\bm{k}} ) \,,\label{eq:chern}
\end{equation}
where $\partial_x$ and $\partial_y$ are partial derivatives along two orthogonal directions in the single-particle BZ, and the integral is over all $\bm{k} \in \mathrm{BZ}$. Here we are only interested in the trivial phase with $C_\pm = 0$. Next, we include the interaction terms and use Lanczos exact diagonalization (ED) to evaluate eigenvalues, eigenstates, and observables of model~\eqref{eq:model} on finite clusters with periodic boundary conditions.

\section{Repulsion-driven charge order}\label{sec:co}

\begin{figure*}[t]
\includegraphics[width=0.95\textwidth]{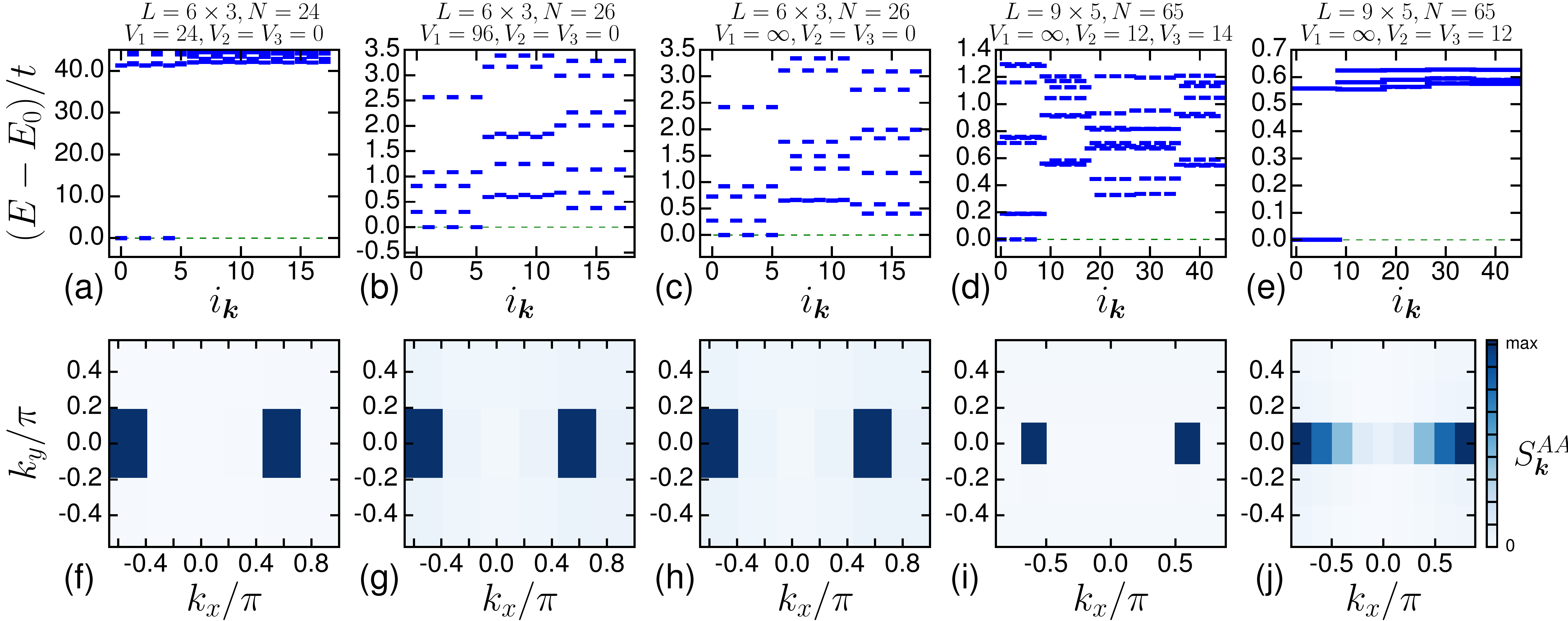}
\caption{(a-e) Many-body energy spectra obtained with ED, with eigenenergies indexed by $i_{\bm{k}}$ running over available momenta in the BZ for each cluster, and (f-j) corresponding GS SSF $S^{AA}_{\bm{k}}$. Each of the GS levels in (e) is multiply degenerate. In all panels, $t=2$, $t_2=1$, $\phi=\pi/3$.}
\label{fig:evnk}
\end{figure*}

For concreteness, from now on we fix $t=2$ and $\phi=\pi/3$, but we have verified that all the results that follow are insensitive to the precise choice of kinetic parameters, as long as one remains in the trivial phase of $\widehat{H}^{\,}_{\textrm{kin}}$. We begin with $V_1\gg t$ and $V_2=V_3=0$. At density $\rho=\rho_{\mathrm{CO}}=2/3$, the repulsion stabilizes charge-order (CO) with the pattern sketched in Fig.~\ref{fig:model}(c). This CO configuration and its two translations by $\bm{a}_1$ and $2\bm{a}_1$ are the only ground states (GSs) in the classical limit $V \gg t,t_2$. We have verified that the system remains ordered for all ratios $t/t_2$ by observing persistent threefold GS degeneracy and sharp peaks in the diagonal components of the GS static structure factor (SSF)
\begin{equation}
 S^{AA}_{\bm{k}} = \left| \sum_{\bm{i}\in A} e^{\mathrm{i} \bm{k}\cdot\bm{i}} (\widehat{n}_{\bm{i}}-\rho) \ket{0} \right|^2 \,,
\end{equation}
where $\ket{0}$ is a state in the GS manifold. This is illustrated in Figs.~\ref{fig:evnk}(a,f). The sharp SSF peaks grow with increasing $V$ and signify CO with ordering vectors $\pm\bm{K}=2\pi/3$. In the remainder of this paper, we will be working exclusively in the limit $V_1 \gg V_2,V_3 \gg t,t_2$. Also, we will only consider clusters of sizes $L_1 \times L_2$ such that the CO is \textit{commensurate}. This CO pattern will be the building block of what follows.

\section{Symmetry-breaking topological order}\label{sec:sbt}

We seek to characterize the ground state that results upon doping the CO with particles. Note that the 36-site cluster shown in Fig.~\ref{fig:model}(c) is the largest one we can access with conventional ED. To access larger systems, we consider the limit $V_1 \to \infty$, in which a large number of configurations are severely penalized energetically and therefore do not participate appreciably in low-energy states. For a system of $N_s = 2 L_1 \times L_2$ sites, the number of NN bonds in the CO is simply $N_s$. In the dilute dopant limit, the number of extra NN bonds due to the presence of dopants is $6 (N - \rho N_s)$. Configurations with a total number of NN bonds greater than $6 N - 3 N_s$ incur an infinite energy cost and are projected out. This projection strategy is customary for the triangular lattice~\cite{Wang2009a}. We verified by comparison of numerical results for original vs projected models that this indeed captures all the essential properties of model~\eqref{eq:model} for systems of up to 36 sites with 26 particles: Figs.~\ref{fig:evnk}(b,g) and (c,h) show that the energy spectrum and GS SSF are qualitatively the same. With this projection, we reach systems of up to $N_s=90$ and $N=66$.

Let us now assume that, upon doping, the CO remains intact, and that dopants reside in the part of the lattice that remains mainly unoccupied by the CO, i.e., the colored part of Fig.~\ref{fig:model}(c). In this scenario, the kinetics of the dopants is effectively governed by the second-NN terms in Eq.~\ref{eq:model}. These terms, when considered alone, constitute the effective two-band model of Refs.~\cite{Venderbos2011,Venderbos2011a,Kourtis2012a,Kourtis2013}. This model has $C_\pm = \pm 1$ and gives rise to FCI states when NN repulsion is included. At a density of $\rho = \rho_{\mathrm{CO}}$ particles per site, these effective Chern bands are empty. To reach a filling fraction $\nu$ of the lower effective Chern band, one needs to dope with an extra density of $\nu$ particles per effective unit cell, or equivalently $\nu$ particles per 6 sites of the original triangular lattice [see Fig.~\ref{fig:model}(c)]. Therefore, at overall densities per site $\rho = \rho_{\mathrm{CO}} + \nu/6$, we expect to obtain a Chern band of dopants with $C_-=-1$ at filling $\nu$, even though the \textit{actual} noninteracting model is topologically trivial with $C_\pm = 0$. If $\nu$ is a FQH fraction and dopants also interact via a finite $V_2$, then a FCI state is likely. Such a composite CO-FCI state is an SBT order that arises purely from building correlations in a topologically trivial noninteracting system.

\begin{figure}[t]
\includegraphics[width=0.99\columnwidth]{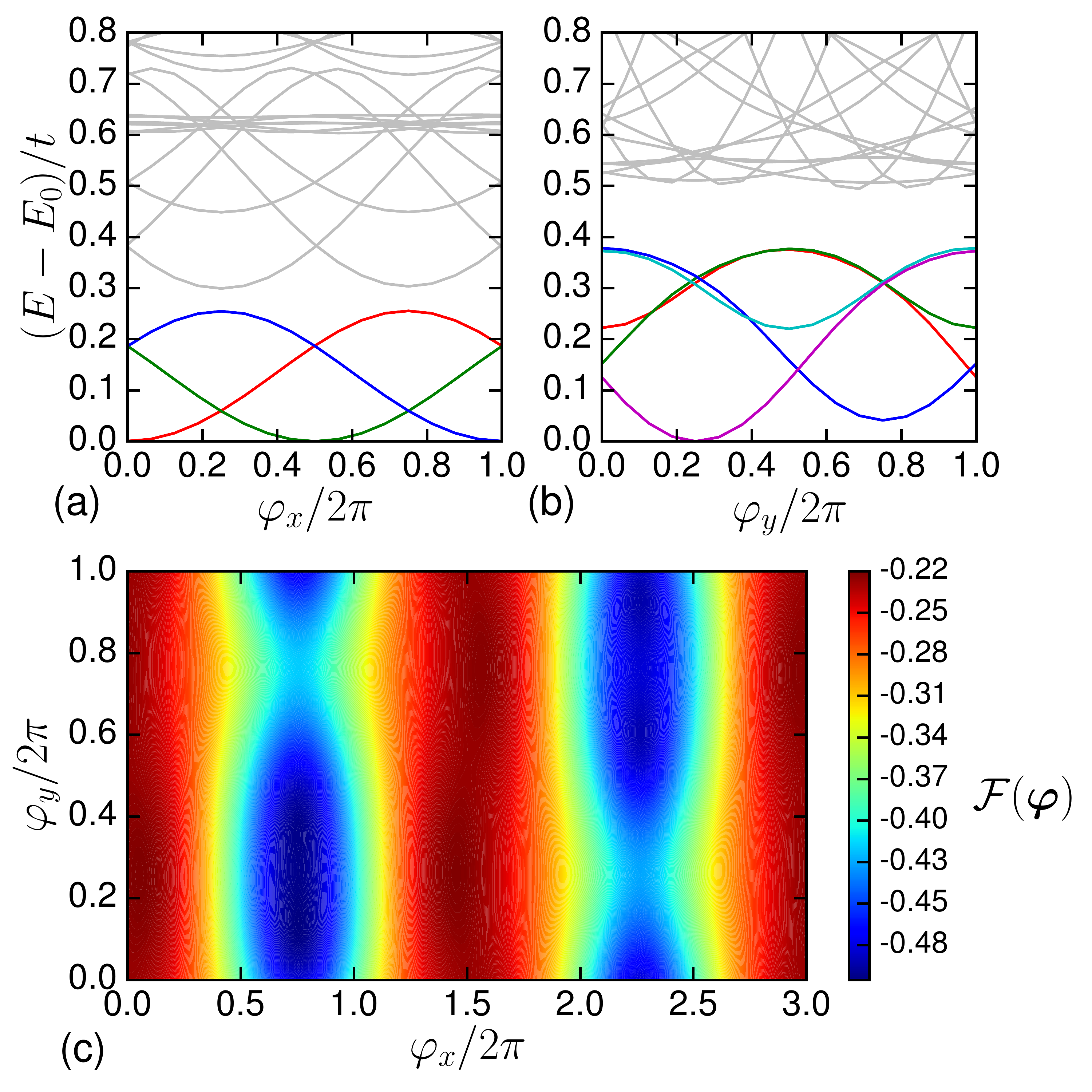}
\caption{(a,b) Spectral flow at (a) $\rho=13/18$ with $V_2=24$, $V_3=25$, (b) $\rho=11/15$ with $V_2=12$, $V_3=14$, and (c) many-body Berry curvature of SBT GS at $\rho=13/18$ with $V_2=12$ and $V_3=14$. All levels shown in (a,b) are triply degenerate. The Berry curvature in (c) is evaluated over a grid of $144\times48$ points and integrates to $1/3$ to accuracy better than $10^{-10}$. In all panels, $L=9\times5$ (90 sites) in the $V_1\to\infty$ limit, with $t=2$, $t_2=1$, $\phi=\pi/3$.}
\label{fig:fci}
\end{figure}

In Fig.~\ref{fig:fci} we present ED results that verify this scenario, showing definitive signatures of FQH-type SBT order. For $\nu=1/3$ ($\rho=13/18$) and $V_2=24$, $V_3=25$ (this choice is explained below) we find a $3\times3$-fold quasi-degenerate GS, where the first factor of 3 is due to the CO and the second is the topological degeneracy of the FCI of dopants~\cite{Kourtis2013}. We insert magnetic fluxes $(\varphi_x,\varphi_y)\equiv\boldsymbol\varphi$ through the handles of the toroidal cluster and upon varying them we observe spectral flow [Fig.~\ref{fig:fci}(a)]: levels exchange place upon insertion of flux $2\pi$, as is typical for FQH states~\cite{Tao1984,Wen1990}, although here they do so in exactly degenerate triplets. We also obtain the Hall conductivity~\cite{Thouless1982,Niu1985a}
\begin{equation}
\sigma^{\,}_{\mathrm{H}} = \frac{e^2}{h} \frac{1}{D} \sum_{n=1}^{D} \int\limits_{0}^{2\pi}\int\limits_{0}^{2\pi} \frac{\mathrm{d}\varphi^{\,}_{x}\,\mathrm{d}\varphi^{\,}_{y}}{4\pi^2} {\cal F}^{\,}_{n}(\boldsymbol\varphi) \,, \label{eq:hall}
\end{equation}
where the sum is over the $D$-fold degenerate GSs and ${\cal F}^{\,}_{n}(\boldsymbol\varphi)$ is the many-body Berry curvature
\begin{equation}
{\cal F}^{\,}_{n}(\boldsymbol\varphi) = 4\pi \mathrm{Im}\! \sum_{n'\not=n}\! \frac{\bra{n} \partial^{\,}_{\varphi^{\,}_{y}} \widehat{H} \ket{n'} \bra{n'} \partial^{\,}_{\varphi^{\,}_{x}} \widehat{H} \ket{n}}{(E^{\,}_{n'}-E^{\,}_{n})^{2}} \,. \label{eq:berry}
\end{equation}
Here $\ket{n}, \ket{n'}$ are ground and excited eigenstates with energies $E_n, E_{n'}$, respectively. Fig.~\ref{fig:fci}(c) shows ${\cal F}^{\,}_{n}$ for one of the quasi-degenerate GSs. Upon integration, we find $\sigma^{\,}_{\mathrm{H}} = \frac13 \frac{e^2}{h}$ with remarkable accuracy. Finally, Figs.~\ref{fig:evnk}(d,i) demonstrate that the CO remains intact, despite the doping and additional interactions. We obtain analogous results at $\rho=11/15$ ($\nu=2/5$): $3\times5$-fold degeneracy, spectral flow [Fig.~\ref{fig:fci}(b)] and $\sigma^{\,}_{\mathrm{H}} = \frac25 \frac{e^2}{h}$. The spread between GS levels in Figs.~\ref{fig:fci}(a,b) is a finite-size effect and disappears in the thermodynamic limit~\cite{Wen1990}. It is pronounced because the \textit{residual} lattice that hosts the dopants is still just $3\times5$ unit cells. Note that we have done no fine-tuning of $\widehat{H}_{\mathrm{kin}}$ to favor SBT order.

\section{Stability against phase separation}\label{sec:ps}

\begin{figure}[t]
\includegraphics[width=0.99\columnwidth]{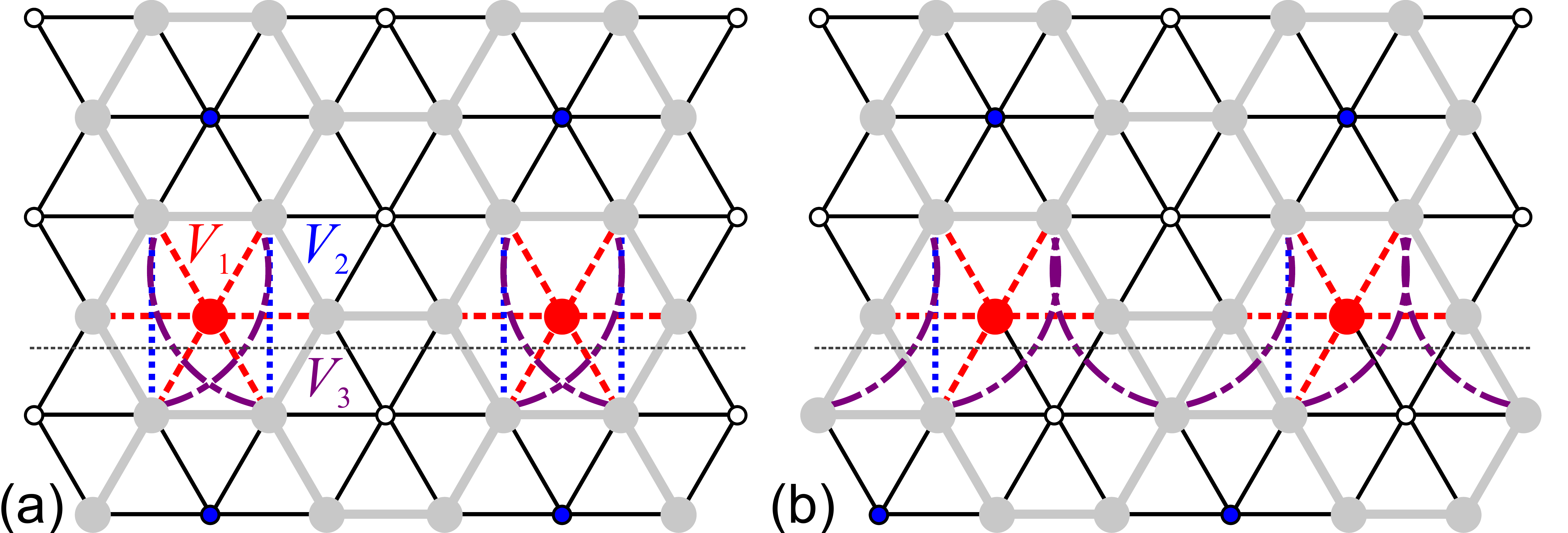}
\caption{Energetics of ``slips'' along domain walls of dopants aligned in a straight line, in the limit $V_1 \gg V_2,V_3 \gg t,t_2$. Sites and NN bonds blocked by the CO are shown in light grey. Dopants are red full circles and their NN bonds are dashed red lines. The thin dotted gray line marks the domain wall. Dotted blue and dash-dotted purple lines denote second- and third-NN bonds across the domain wall, respectively. In (b) all particles below the domain wall are translated by one site to the left with respect to (a). The numbers of first-, second-, and third-NN bonds \textit{per dopant} across the domain wall are (a) 4, 2, and 2, and (b) 4, 1, and 3, respectively.}
\label{fig:ps}
\end{figure}

When interactions are restricted to NN range ($V_2=V_3=0$) and $V_1 \gg t$, it is expected that model~\eqref{eq:model} phase-separates for $\rho > 2/3$~\cite{Wessel2005,Boninsegni2005,Heidarian2005,Melko2005,Burkov2005}: it is favorable for dopants to align themselves in a straight line, as shown in Fig.~\ref{fig:ps}, allowing for the formation of domain walls, and ``slips'' along these walls incur zero energy cost. This, in turn, allows the dopants to hop across domain walls and reduce the kinetic energy of the system. As illustrated in Fig.~\ref{fig:ps}, a finite $V_2>0$ will favor domain walls, as now there is also a potential energy gain in their formation: there is only 1 second-NN bond per dopant in Fig.~\ref{fig:ps}(b) instead of the 2 second-NN bonds per dopant in Fig.~\ref{fig:ps}(a). In contrast, a finite $V_3>0$ hinders the formation of domain walls: there are 3 third-NN bonds per dopant in Fig.~\ref{fig:ps}(b) instead of only 2 such bonds per dopant in Fig.~\ref{fig:ps}(a). By this counting, when $V_1 \gg V_2,V_3 \gg t,t_2$, it is seen that ``slips'' along domain walls, which make the formation of domains energetically favorable, are suppressed when $V_3>V_2$.

\begin{figure}[b]
\includegraphics[width=0.99\columnwidth]{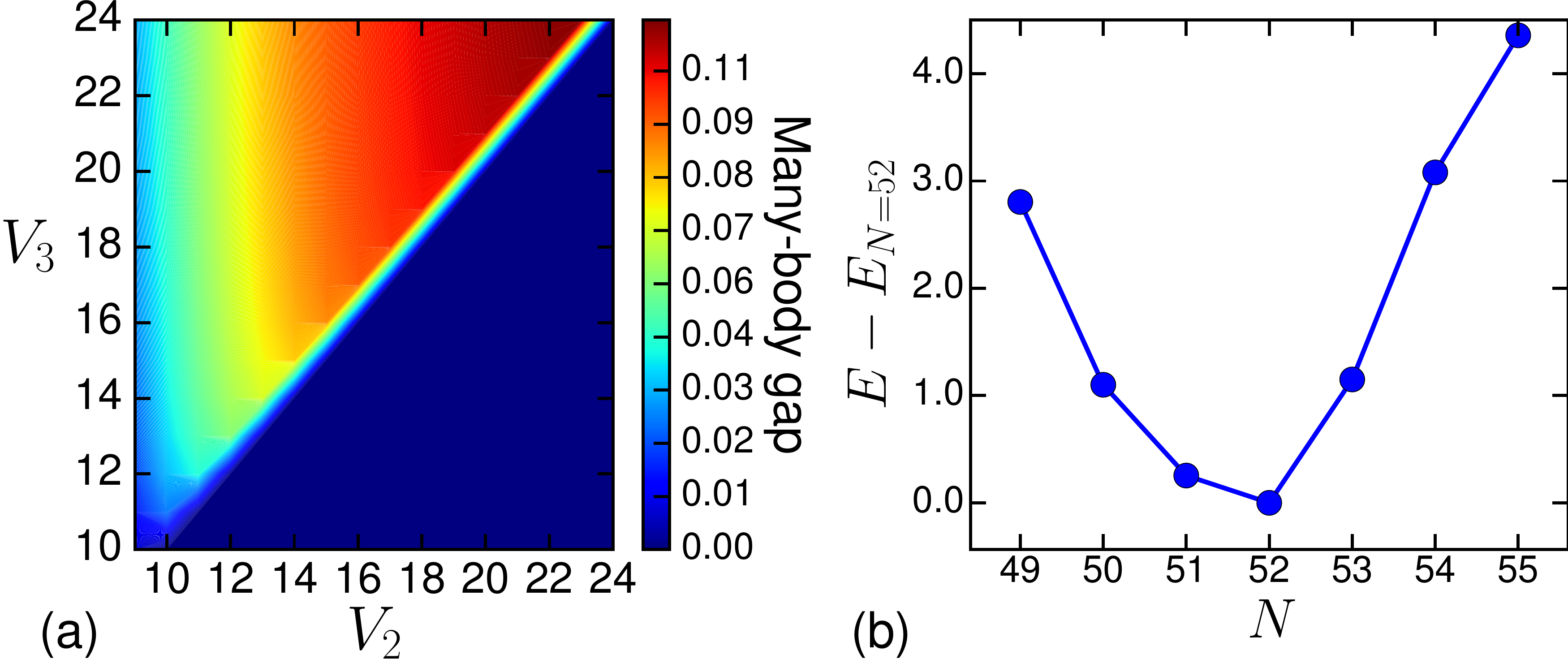}
\caption{(a) $V_2$-$V_3$ phase diagram at $\rho=13/18$. Color scale represents the \textit{minimum} gap in $\boldsymbol\varphi$-space. (b) $E_0$ vs $N$ is convex with minimum at $\rho=13/18$ for $\mu=-120$. In both panels, $L=9\times4$ (72 sites) in the $V_1\to\infty$ limit, $t=2$, $t_2=1$, $\phi=\pi/3$.}
\label{fig:phdiag}
\end{figure}

Our ED results corroborate this heuristic [see Fig.~\ref{fig:phdiag}(a)]. First, for densities up to $\rho=3/4$, $S^{AA}_{\bm{k}}$ shows sharp peaks at precisely $\pm\bm{K}$ whenever $V_3>V_2$ [see, e.g., Fig.~\ref{fig:evnk}(i)]. This holds irrespective of the value of $t_2$ (longer-range hoppings further counteract domain wall formation~\cite{Dang2008}). In contrast, for $V_3 \le V_2$ we find an extensive number of GSs and a SSF that develops broader features, with maxima that are not anymore at $\pm \bm{K}$ [Figs.~\ref{fig:evnk}(e,j)]. Second, at $\rho=13/18$ ($\nu=1/3$), for clusters of $N_s=72,90$ sites, we find a gapped and non-extensive manifold of 9 quasi-degenerate SBT GSs in a wide range of $V_3>V_2$, that is robust against small variations of all other parameters of the model. The $V_2$-$V_3$ phase diagram is presented in Fig.~\ref{fig:phdiag}(a). Finally, the total energy as a function of $N$ for the 72-site cluster, shown in Fig.~\ref{fig:phdiag}(b), is convex with a minimum at $\rho=13/18$ for a range of values of $\mu$, indicating a thermodynamically stable phase.

\section{Summary and outlook}\label{sec:outlook}

We have presented a viable route towards SBT ordered states, which arise by introducing strong repulsion in topologically trivial phases of simple noninteracting lattice models that break TR symmetry. We have demonstrated this by constructing a minimal model of interacting spinless fermions, studying it with exact diagonalization in the strong NN-repulsion limit, and discovering compelling evidence for the first instances of \textit{fermionic} FQH-type topological order with fractionally quantized Hall conductivity $\sigma_{\mathrm{H}}=1/3,2/5$ in units of $e^2/h$ in a minimal model of two $C_\pm=0$ bands. We investigated the effect of second- and third-NN interactions $V_2$ and $V_3$, and found that SBT order is robust against phase separation for $V_2 < V_3$.

Further numerical and analytical handles for FQH-like SBT ordered states can be provided by density-matrix renormalization group~\cite{Liu2013,Grushin2015,Gong2015,Zhu2016b} and effective theories~\cite{Zhang1992,Sohal2017}, respectively. Theoretical explorations can be guided by physical systems that fulfill the requirements for the formation of such correlated states. For example, AgNiO${}_2$~\cite{Coldea2014,F??vrier2015} is a quasi two-dimensional compound that incorporates (i) strong interactions beyond onsite repulsion, (ii) coexistence of charge order and itinerant carriers, and (iii) noncollinear magnetic order that could lead to nontrivial TR-breaking flux arrangement upon application of an external field. Alternatively, spin-orbit coupled correlated oxides~\cite{Witczak-Krempa2014} may offer other viable settings for the emergence of FQH-type SBT order or its plausible TR-symmetric generalization.

\begin{acknowledgments}
The author is grateful to M.~Daghofer for many illuminating discussions and comments on this work, and to C.~Chamon and A.~Grushin for critical reading of the manuscript. This work was partially supported by the Boston University Center for Non-Equilibrium Systems and Computation.
\end{acknowledgments}

\bibliographystyle{apsrev4-1}

\end{document}